\documentclass[aps,twocolumn,showpacs,preprintnumbers,amsmath,amssymb,superscriptaddress,floatfix,nofootinbib]{revtex4}

\usepackage{graphicx}
\usepackage{epsfig}
\usepackage{epstopdf}
\usepackage{hyperref}
\usepackage{amsmath}
\usepackage{amsfonts}
\usepackage{amssymb}
\usepackage{multirow}
\usepackage[dvipsnames]{xcolor}
\usepackage{ulem}
\usepackage{longtable}
\usepackage{array}

\newcommand{\Slash}[1]{\ooalign{\hfil/\hfil\crcr$#1$}}

\newcommand{\be}{\begin{equation}}
\newcommand{\ee}{\end{equation}}
\newcommand{\bea}{\begin{eqnarray}}
\newcommand{\eea}{\end{eqnarray}}
\newcommand{\bean}{\begin{eqnarray*}}
\newcommand{\eean}{\end{eqnarray*}}

\begin{document}

\title{Exploratory study of possible resonances in the heavy meson - heavy baryon coupled-channel interactions}

\author{Chao-Wei Shen} \email{shencw@itp.ac.cn}
\affiliation{Key Laboratory of Theoretical Physics, Institute of Theoretical Physics,
Chinese Academy of Sciences, Beijing 100190,China}
\affiliation{University of Chinese Academy of Sciences (UCAS), Beijing 100049, China}

\author{Deborah R\"onchen} \email{roenchen@hiskp.uni-bonn.de}
\affiliation{Helmholtz-Institut f\"ur Strahlen- und Kernphysik (Theorie) and Bethe Center for Theoretical Physics,
Universit\"at Bonn, D-53115 Bonn, Germany}

\author{Ulf-G.~Mei\ss ner}
\affiliation{Helmholtz-Institut f\"ur Strahlen- und Kernphysik (Theorie) and Bethe Center for Theoretical Physics,
Universit\"at Bonn, D-53115 Bonn, Germany}

\author{Bing-Song Zou}
\affiliation{Key Laboratory of Theoretical Physics, Institute of Theoretical Physics,
Chinese Academy of Sciences, Beijing 100190,China}
\affiliation{University of Chinese Academy of Sciences (UCAS), Beijing 100049, China}


\begin{abstract}
We use a unitary coupled-channel model to study the $\bar{D} \Lambda_c - \bar{D} \Sigma_c$ interactions.
In our calculation, SU(3) flavor symmetry is applied to determine the coupling constants.
Several resonant and bound states with different spin and parity are dynamically generated in the mass range of the recently observed pentaquarks.
The approach is also extended to the hidden beauty sector to study the $B \Lambda_b - B \Sigma_b$ interactions.
As the $b$-quark mass is heavier than the $c$-quark mass, there are more resonances observed
for the $B \Lambda_b - B \Sigma_b$ interactions and they are more tightly bound.
\end{abstract}

\maketitle

\section{Introduction}

In 2015, the LHCb Collaboration observed two exotic resonant states
in the $J/\psi p$ invariant mass distribution from $\Lambda^0_b \to J/\psi p K^-$ decays~\cite{Aaij:2015tga}.
Following this discovery, those structures were addressed in numerous  model calculations and explained, e.g., as hidden charm pentaquark-like states with opposite parity~\cite{Aaij:2015tga,Shen:2016tzq,Chen:2015loa,Chen:2015moa,
Roca:2015dva,Mironov:2015ica,He:2015cea,Huang:2015uda,Burns:2015dwa,Lu:2016nnt,He:2016pfa,Lin:2017mtz,Guo:2017jvc,Guo:2015umn,Meissner:2015mza,
Liu:2015fea,Lebed:2015tna,Wang:2015epa,Scoccola:2015nia,Zhu:2015bba,Shimizu:2016rrd,Yamaguchi:2016ote,Wang:2016dzu}.
They are now listed as $P_c(4380)^+$ and $P_c(4450)^+$  pentaquarks in the $Review \ of \ Particle \ Physics$~\cite{Olive:2016xmw}.
Among these theoretical works, a popular explanation is that these two resonances are
bound $\bar{D} \Sigma_c^*(2520)$ and $\bar{D}^* \Sigma_c(2455)$ molecular states~\cite{Shen:2016tzq,Chen:2015loa,Chen:2015moa,Roca:2015dva,Mironov:2015ica,He:2015cea,Huang:2015uda,Burns:2015dwa,Lu:2016nnt,He:2016pfa,Lin:2017mtz,Guo:2017jvc}.
The study of exotic hadrons with more than three constituent quarks
has been an important issue in hadron physics for a long time.
Since 2010, five years prior to the experimental observation, such pentaquark-like states with hidden charm
have already been predicted by different groups~\cite{Wu:2010jy,Wu:2010vk,Wang:2011rga,Yang:2011wz,Yuan:2012wz,Wu:2012md,Xiao:2013yca,Uchino:2015uha}.
The discovery of the $P_c(4380)$ and $P_c(4450)$ thus adds new momentum to the effort to study the baryon spectrum in the 4~GeV energy range, aiming at the investigation of other pentaquark candidates and, eventually, even at a complete picture of the pentaquark spectrum.


The preferred spin of $P_c(4380)$ and $P_c(4450)$ is one having spin $J=3/2$ and the other 5/2, no spin 1/2 state has been observed yet.
In Refs.~\cite{Wu:2010jy,Wu:2010vk}, Wu {\it et~al.} used a coupled-channel unitary approach with the local hidden gauge formalism
to calculate the interactions of $\bar{D} \Lambda_c - \bar{D} \Sigma_c$ and $\bar{D}^* \Lambda_c - \bar{D}^* \Sigma_c$. In this analysis only vector meson exchange is considered.
 The lowest $\bar{D}^* \Sigma_c$ molecular state with $J^P=3/2^-$ is predicted to be around 4412~MeV, which lies in the middle of the two $P_c$ states observed by the LHCb Collaboration.
In the $\bar{D} \Lambda_c - \bar{D} \Sigma_c$ interaction, the lowest hidden charm resonance of $J^P=1/2^-$ is predicted to have a  mass of 4261~MeV. In these earliest predictions, only S-wave molecules are considered.   It should be noted, however,  that this study employs some
approximations that have recently been scrutinized in Ref.~\cite{Gulmez:2016scm}.

In the present work, we aim at a more complete picture of the resonance spectrum in the $\bar{D} \Lambda_c - \bar{D} \Sigma_c$ system focusing on possible states in lower and higher partial waves.
To this goal we extend the J\"ulich-Bonn dynamical coupled-channel (J\"uBo DCC) framework~\cite{Ronchen:2012eg}, a unitary meson-baryon exchange model, to the hidden charm sector. In dynamical coupled-channel approaches different reactions are analysed simultaneously and partial waves of higher order are taken into account. Moreover, theoretical constraints of the $S$-matrix like unitarity and analyticity are respected. They provide thus an ideal tool to extract resonance parameters such as pole positions and residues from experimental data.

The present work should be regarded as an exploratory study, as only two channels, $\bar{D} \Lambda_c$ and  $\bar{D} \Sigma_c$, are considered. Nevertheless, we are able to explore the possibility of dynamically generated poles in different partial waves in the 4~GeV energy range. In subsequent studies we plan to include also $\bar D^*Y_c$ and lighter meson-baryon channels, which will provide a more comprehensive picture.

We will proceed likewise in extending the formalism to the hidden beauty sector and search for dynamically generated poles in the $B \Lambda_b - B \Sigma_b$ system.


This article is organized as follows. In Sect.~\ref{sec:formalism}, we present
the theoretical framework of our calculation. In Sect.~\ref{sec:results},
the numerical results for the $\bar{D} \Lambda_c - \bar{D} \Sigma_c$ and $B \Lambda_b - B \Sigma_b$ interactions are discussed,
followed by a brief summary of our findings.

\section{Theoretical framework} \label{sec:formalism}

The J\"uBo DCC model has been developed over the years and was originally constructed to describe elastic and inelastic $\pi N$ scattering. For a detailed description of the approach see Refs.~\cite{Ronchen:2012eg,Doring:2010ap} and references therein. More recently, the framework was extended to pseudoscalar meson photoproduction~\cite{Ronchen:2014cna,Ronchen:2015vfa,Huang:2011as}. The spectrum of $N^*$ and $\Delta^*$ resonances is extracted by fitting the free parameters inherent to the approach to experimental data from pion- and photon-induced hadronic reactions. Analyticity and two-body unitarity are manifestly implemented, while three-body unitarity is approximately fulfilled. Left-hand cuts and the correct structure of complex branch points are included. This ensures the well defined determination of resonance parameters in terms of poles in the complex energy plane of the scattering amplitude and the corresponding residues. The analytic properties of the scattering amplitude are discussed in detail in Ref.~\cite{Doring:2009yv}.

The scattering equation that describes the interaction of a baryon and a meson reads
\begin{eqnarray}
T_{\mu \nu}(p^{\prime\prime},p^\prime,z) &=& V_{\mu \nu}(p^{\prime\prime},p^\prime,z)
             + \sum_{\kappa} \int_0^\infty \!\!\!{d}p\, p^2 V_{\mu \kappa}(p^{\prime\prime},p,z)  \nonumber \\
             && \times G_\kappa(p,z) T_{\kappa \nu}(p,p^\prime,z). \label{eq:Tmat}
\end{eqnarray}
Eq.~(\ref{eq:Tmat}) is formulated in partial-wave basis and $z$ is the scattering energy in the center-of-mass system, $p^{\prime\prime} \equiv \Vert \vec{p^{\prime\prime}} \Vert$
and $p^\prime \equiv \Vert \vec{p^\prime} \Vert$ represent the out-going and in-coming three-momentum that may be on- or off-shell, while $\mu$, $\nu$ and $\kappa$ are channel indices. The propagator $G_\kappa$ for channels with stable particles is given by
\begin{equation}
G_\kappa(p,z) = \frac{1}{z - E_a(p) - E_b(p) + i \epsilon}, \label{eq:prop}
\end{equation}
with $E_a(p)=\sqrt{m_a^2+p^2}$ and $E_b(p)=\sqrt{m_b^2+p^2}$  the on-mass-shell energies of the intermediate particles $a$ and $b$ in channel $\kappa$ with the respective masses $m_a$ and $m_b$.
The J\"uBo approach also includes the three-body $\pi\pi N$ channel, effectively parameterized as $\rho N$, $\sigma N$, and $\pi\Delta$, see, e.g. Ref.~\cite{Schutz:1998jx}.

The scattering potential $V_{\mu\nu}$ in Eq.~(\ref{eq:Tmat}) consists of $s$-channel processes that account for ``genuine'' resonance states, and $t$- and $u$-channel exchanges of known mesons and baryons that constitute the non-resonant part of the amplitude. In addition, contact interaction terms are included~\cite{Ronchen:2015vfa}. Note that while the $t$- and $u$-channel exchanges are often referred to as the non-pole part of the $T$-matrix or the background, the dynamical generation of poles through the unitarization of Eq.~(\ref{eq:Tmat}) is possible.

The potential $V_{\mu\nu}$ is derived from an effective chiral Lagrangian using time-ordered perturbation theory (TOPT). Explicit expressions for all exchange processes included in the analysis of elastic and inelastic $\pi N$ scattering and the corresponding Lagrangians can be found in Ref.~\cite{Ronchen:2012eg}. Explicit expressions for $s$-channel diagram are given in Ref.~\cite{Doring:2010ap}.

In the present study, we extend this formalism to the hidden charm and hidden beauty sectors, {\sl i.e.}, we solve Eq.~(\ref{eq:Tmat}) for $\mu,\nu=\bar{D} \Lambda_c$, $\bar{D} \Sigma_c$ and $B\Lambda_b$, $B \Sigma_b$. As mentioned above, the inclusion of further channels such as $\bar D^*Y_c$ or $\pi N$ and $\eta N$ is postponed to a subsequent analysis.

\begin{figure}[htbp]
\centering
\includegraphics[scale=0.7]{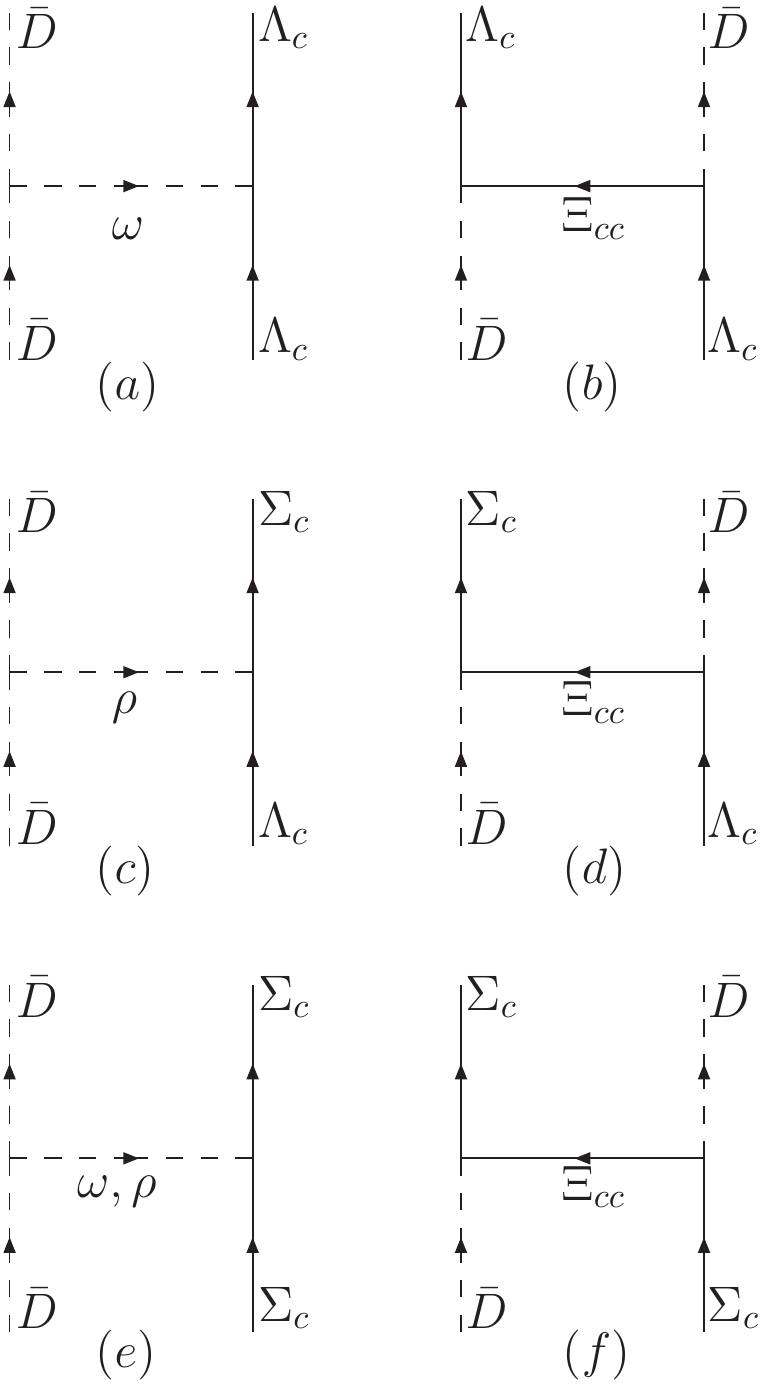}
\caption{Feynman diagrams for $\bar{D} \Lambda_c \to \bar{D} \Lambda_c$
via (a) $t$-channel $\omega$ exchange, (b) $u$-channel $\Xi_{cc}$ exchange;
$\bar{D} \Lambda_c \to \bar{D} \Sigma_c$
via (c) $t$-channel $\rho$ exchange, (d) $u$-channel $\Xi_{cc}$ exchange;
$\bar{D} \Sigma_c \to \bar{D} \Sigma_c$
via (e) $t$-channel $\omega, \rho$ exchange, (f) $u$-channel $\Xi_{cc}$ exchange.
\label{Fig:feyndiag}}
\end{figure}

We consider only $t$-channel meson and $u$-channel baryon exchanges in the current analysis, the corresponding Feynman diagrams are shown in Fig.~\ref{Fig:feyndiag}. Genuine resonances in form of $s$-channel processes are not included, the same applies to contact interaction terms.

In principle, one could also include a $\phi$ $t$-channel exchange in addition to the diagrams shown in Fig.~\ref{Fig:feyndiag}.
However, since $\phi$ is a pure $s\bar{s}$ particle,
its coupling to heavy mesons like $\bar{D}$ or $B$ should be negligible.
Similar arguments apply in case of the $f_0(980)$. Questions about its inner structure are not yet fully resolved,
e.g., the quark content of the $f_0(980)$ is frequently regarded to have a substantial $K\bar{K}$ molecule component instead of being a pure $s\bar{s}$ state~\cite{Olive:2016xmw,Baru:2003qq,Kroll:2016mbt}.
Its coupling to $\bar{D}$ or $B$ mesons is also small and we do not include the $f_0(980)$ in our calculation.
Thus, in the $t$-channel only $\rho$ and $\omega$ exchange are considered.
In the $u$-channel we take into account the exchange of a doubly-charmed $\Xi_{cc}$ baryon. As the mass of the $\Xi_{cc}$ is very heavy
its contribution should be small.
It turns out that in the present calculation vector meson exchange is indeed the dominant contribution.

For the $B \Lambda_b - B \Sigma_b$ interactions, we also take into account $t$-channel $\rho$ and $\omega$ exchange.
As there is no strong evidence for the existence of a doubly-beauty baryon until now, no $u$-channel baryon exchanges are included.

In the following we briefly describe the construction of the scattering potential $V_{\mu\nu}$ for the channels considered here. For a more detailed account of all components entering the J\"ulich-Bonn model, the reader is referred to Ref.~\cite{Ronchen:2012eg} and references therein.

The effective Lagrangians for the interactions of two pseudoscalar mesons with one vector meson, two baryons with one vector meson and two baryons with one pseudoscalar meson are~\cite{Ronchen:2012eg}
\begin{eqnarray}
{\cal L}_{PPV} &=& g_{PPV} \phi_P(x) \partial_{\mu} \phi_P(x) \phi_V^{\mu}(x), \nonumber \\
{\cal L}_{BBV} &=& g_{BBV} \bar{\psi}_B(x) (\gamma^\mu - \frac{\kappa}{2m_B} \sigma^{\mu \nu} \partial_{\nu}) \phi_V^\mu(x) \psi_B(x), \nonumber \\
{\cal L}_{BBP} &=& \frac{g_{BBP}}{m_P} \bar{\psi}_B(x) \gamma^5 \gamma^\mu \partial_{\mu} \phi_P(x) \psi_B(x), \label{eq:lag}
\end{eqnarray}
where $P$, $V$ and $B$ denote the pseudoscalar meson, vector meson and octet baryon, respectively.

The exchange potentials $V_{t,u}$ in Eq.~(\ref{eq:Tmat}) consist of the pseudo-potential $\mathcal{V}_{t,u}$ multiplied by a kinematic normalization factor $N$, form factors $F(q)$ and an isospin factor IF:
\begin{equation}
V_{t,u} (p_1,p_2,p_3,p_4) = N\,F_a(q)F_b(q) \,{\rm IF} \, {\cal V}_{t,u}\,. \label{eq:amp}
\end{equation}

The pseudo-potentials $\mathcal{V}_{t,u}$ applied in the present study are
\begin{eqnarray}
{\cal V}_{t} &=& g_a \bar{u}(\vec p_3,\lambda_3) \bigg(\frac{g_b \gamma^{\mu} - i \frac{f_b}{2m_B} \sigma^{\mu \nu} q_{\nu}}{z - \omega_q - E_3 - \omega_2} \nonumber \\
            && + \frac{g_b \gamma^{\mu} - i \frac{f_b}{2m_B} \sigma^{\mu \nu} \tilde{q}_{\nu}}{z - \omega_q - E_1 - \omega_4} \bigg)
            u(\vec p_1,\lambda_1) \frac{(p_2+p_4)_{\mu}}{2\omega_q}, \nonumber \\
{\cal V}_{u} &=& \frac{g_a g_b}{m_P^2} \bar{u}(\vec p_3,\lambda_3) \gamma^5 \frac{\Slash p_2}{2E_q} \bigg(\frac{\Slash q + m_{ex}}{z-E_q-\omega_2-\omega_4} \nonumber \\
            && + \frac{\tilde{\Slash q} +m_{ex}}{z-E_q-E_1-E_3}\bigg) \gamma^5 \Slash p_4 u(\vec p_1,\lambda_1). \label{eq:pot}
\end{eqnarray}
Here, the indices 1 and 2 (3 and 4) represent the in-coming (out-going) baryon and meson.
$E_i$ and $\omega_i$ are the on-shell energies for the baryon and the meson, and $q$ is the exchange particle's momentum. We have $\vec q=\vec p_1-\vec p_3$ for $t$-channel and $\vec q=\vec p_1-\vec p_4$ for $u$-channel.
In the TOPT framework used in this study, the four momentum $q$ has $q^0=E_q(\omega_q)$ for baryon (meson) exchange in the first time ordering while $\tilde{q}$ denotes the second time ordering with  $\tilde{q}^0=-E_q(-\omega_q)$ for baryon (meson) exchange.
The three momentum of $q$ and $\tilde{q}$ is the same, denoted as  $\vec{q}$.

We apply SU(3) flavor symmetry to relate the coupling constants $g_a$, $g_b$ at the different vertices to known couplings of other mesons and baryons. This symmetry is, of course, strongly broken by the large differences in the physical masses of, e.g., the nucleon and the $\Sigma_c$, but also by the form factors of the vertices.
Since in our calculation the involved quarks are only $u$, $d$ and $c$ or $u$, $d$ and $b$,
we do not need to derive these relations under SU(4) symmetry.
Assuming that the $s$, $c$ and $b$ quarks can all be regarded  as the same heavy quark compared to $u$ and $d$, we can directly adopt the SU(3) relations listed in Ref.~\cite{Ronchen:2012eg} for reactions involving hidden strangeness and apply them to the hidden charm and beauty reactions of this study.

The values for $g_a$, $g_b$ appearing in the diagrams of Fig.~\ref{Fig:feyndiag} are given by
\begin{eqnarray}
g_{DD\omega} &=& g_{BB\omega} = \frac{1}{2}g_{\pi\pi\rho}~, \nonumber \\
g_{DD\rho} &=& g_{BB\rho} = \frac{1}{2}g_{\pi\pi\rho}~, \nonumber \\
g_{\Lambda_c \Lambda_c \omega} &=& g_{\Lambda_b \Lambda_b \omega} = \frac{2}{3} g_{NN\rho} (5\alpha_{BBV}-2)~, \nonumber \\
g_{\Sigma_c \Lambda_c \rho} &=& g_{\Sigma_b \Lambda_b \rho} = \frac{2}{\sqrt{3}} g_{NN\rho} (1-\alpha_{BBV})~, \nonumber \\
g_{\Sigma_c \Sigma_c \omega} &=& g_{\Sigma_b \Sigma_b \omega} = 2 g_{NN\rho} \alpha_{BBV}~, \nonumber \\
g_{\Sigma_c \Sigma_c \rho} &=& g_{\Sigma_b \Sigma_b \rho} = 2 g_{NN\rho} \alpha_{BBV}~, \nonumber \\
f_{\Lambda_c \Lambda_c \omega} &=& f_{\Lambda_b \Lambda_b \omega} = \frac{5}{6} f_{NN\omega} - \frac{1}{2} f_{NN\rho}~, \nonumber \\
f_{\Sigma_c \Lambda_c \rho} &=& f_{\Sigma_b \Lambda_b \rho} = -\frac{1}{2\sqrt{3}} f_{NN\omega} + \frac{\sqrt{3}}{2} f_{NN\rho}~, \nonumber \\
f_{\Sigma_c \Sigma_c \omega} &=& f_{\Sigma_b \Sigma_b \omega} = \frac{1}{2} f_{NN\omega} + \frac{1}{2} f_{NN\rho}~, \nonumber \\
f_{\Sigma_c \Sigma_c \rho} &=& f_{\Sigma_b \Sigma_b \rho} = \frac{1}{2} f_{NN\omega} + \frac{1}{2} f_{NN\rho}~, \nonumber \\
g_{\Xi_{cc} \Sigma_c D} &=& -g_{NN\pi}~, \nonumber \\
g_{\Xi_{cc} \Lambda_c D} &=& \frac{1}{\sqrt{3}} g_{NN\pi} (4\alpha_{BBP}-1)\,, \label{eq:coup}
\end{eqnarray}
with $g_{\pi\pi\rho}=6.04$, $g_{NN\rho}=3.25$, $\alpha_{BBV}=1.15$, $f_{NN\rho}=g_{NN\rho}\kappa_{\rho}=19.825$, $f_{NN\omega}=0$, $g_{NN\pi}=0.989$ and $\alpha_{BBP}=0.4$.
Note that the $B$ in $g_{BB\omega}$ and $g_{BB\rho}$ refers to the $B$ meson, while the $B$ in $\alpha_{BBV}$ represents an octet baryon.

At each vertex, an off-shell form factor as shown in Eq.~(\ref{eq:FF}) is used,
\begin{equation}
F(q)=\left(\frac{\Lambda^2-m_{ex}^2}{\Lambda^2 + {\vec{q}}^{\;2}}\right)^n \,,
\label{eq:FF}
\end{equation}
where $m_{ex}$, $\vec q$ and $\Lambda$ are the exchange particle's mass, three momentum and cut-off parameter.
For the $t$-channel $\rho$ and $\omega$ exchange, we use a dipole form factor, i.e. $n=2$, while a monopole form factor is applied for $u$-channel $\Xi_{cc}$ exchange, i.e. $n=1$. As usual in the J\"uBo model, different powers for the form factors are applied to ensure the convergence of the integral over the off-shell momenta in the scattering equation.

In the J\"ulich-Bonn approach, the cut-off values $\Lambda$ are free parameters that are fitted to data. As there is no data available for the reactions considered in this study, we use the values determined in Ref.~\cite{Ronchen:2012eg} in an analysis of the reactions $\pi N\to\pi N$, $\eta N$ and $KY$. Two different fits were performed in Ref.~\cite{Ronchen:2012eg}, fit A and B, starting from two different scenarios in the fit parameter space. In the present study the cut-off parameters are set to the values of the corresponding strange hadron vertices from Ref.~\cite{Ronchen:2012eg}, e.g., $\Lambda_{D\Xi_{cc}\Lambda_c}=\Lambda_{K\Xi\Lambda}$.
The cut-off values for all vertices are listed in table~\ref{table:cutoff}.

The calculations are performed for both sets of parameters A and B. This allows to get a rough estimate of the dependence of the results on the cut-off parameters.

\begin{table}[htbp]
    \begin{center}
\caption{\label{table:cutoff}
    Cut-off parameters $\Lambda$ applied in the calculations.}
        \renewcommand{\arraystretch}{1.30}
    \begin{tabular}{c c c c}
        \hline
        \hline
        Vertex & Exchanged  & \multicolumn{2}{c}{$\Lambda$ [MeV]} \\
               & particle & $\quad$ A $\quad$ & $\quad$ B $\quad$\\
        \hline
        $D \omega D$ or $B \omega B$ & $\omega$ & 1310 & 1430 \\
        \hline
        $D \rho D$ or $B \rho B$ & $\rho$ & 3140 & 2580 \\
        \hline
        $\Lambda_c \omega \Lambda_c$ or $\Lambda_b \omega \Lambda_b$ & $\omega$ & 1100 & 1100 \\
        \hline
        $\Lambda_c \rho \Sigma_c$ or $\Lambda_b \rho \Sigma_b$ & $\rho$ & 1710 & 1750 \\
        \hline
        $\Sigma_c \omega \Sigma_c$ or $\Sigma_b \omega \Sigma_b$ & $\omega$ & 1940 & 1940 \\
        \hline
        $\Sigma_c \rho \Sigma_c$ or $\Sigma_b \rho \Sigma_b$ & $\rho$ & 1480 & 1480 \\
        \hline
        $D \Xi_{cc} \Lambda_c$ & $\Xi_{cc}$ & 1670 & 1670 \\
        \hline
        $D \Xi_{cc} \Sigma_c$ & $\Xi_{cc}$ & 2340 & 2340 \\
        \hline
        \hline
    \end{tabular}
    \end{center}
\end{table}

The kinematic normalization factor $N$ in Eq.~(\ref{eq:amp}) is of the form
\begin{eqnarray}
N = \frac{1}{(2 \pi)^3} \frac{1}{2 \sqrt{\omega_2 \omega_4}}, \label{eq:nor}
\end{eqnarray}
and the isospin factors IF for each exchange process can be found in table~\ref{table:if}.
Note that the baryon-first convention is applied throughout the calculation.

\begin{table}[htbp]
    \begin{center}
\caption{\label{table:if}
    Isospin factors (IF) for isospin I=1/2 and 3/2.}
   \renewcommand{\arraystretch}{1.30}
    \begin{tabular}{c c c c}
        \hline
        \hline
        \ Process \ & Exchanged & \ IF($\frac{1}{2}$) \ & \ IF($\frac{3}{2}$) \ \\
        & particle &  &  \\
        \hline
        $\bar{D} \Lambda_c \to \bar{D} \Lambda_c$ & $\omega$ & 1 & 0 \\
        & $\Xi_{cc}$ & 1 & 0 \\
        \hline
        $\bar{D} \Lambda_c \to \bar{D} \Sigma_c$ & $\rho$ & $-\sqrt{3}$ & 0 \\
        & $\Xi_{cc}$ & $\sqrt{3}$ & 0 \\
        \hline
        $\bar{D} \Sigma_c \to \bar{D} \Sigma_c$ & $\omega$ & 1 & 1 \\
        & $\rho$ & 2 & $-$1 \\
        & $\Xi_{cc}$ & $-$1 & 2 \\
        \hline
        $B \Lambda_b \to B \Lambda_b$ & $\omega$ & 1 & 0 \\
        \hline
        $B \Lambda_b \to B \Sigma_b$ & $\rho$ & $-\sqrt{3}$ & 0 \\
        \hline
        $B \Sigma_b \to B \Sigma_b$ & $\omega$ & 1 & 1 \\
        & $\rho$ & 2 & $-$1 \\
        \hline
        \hline
    \end{tabular}
    \end{center}
\end{table}

\section{Results and discussions} \label{sec:results}

\subsection{$\bar{D} \Lambda_c - \bar{D} \Sigma_c$ interactions}

\begin{figure}[htbp]
\centering
\includegraphics[scale=1.3]{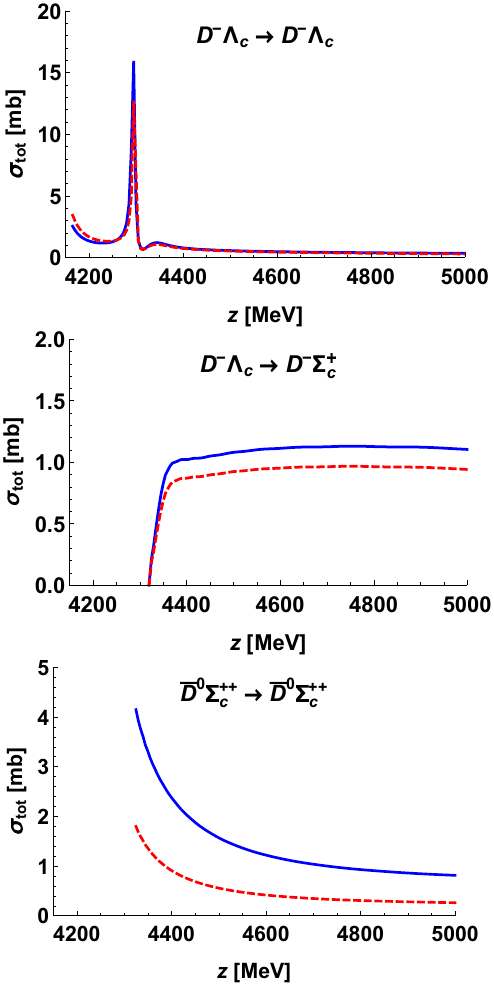}
\caption{Total cross sections of $\bar{D} \Lambda_c \to \bar{D} \Lambda_c$,
$\bar{D} \Lambda_c \to \bar{D} \Sigma_c$ and $\bar{D} \Sigma_c \to \bar{D} \Sigma_c$
using the values of the cut-offs from fit A (blue solid lines) and fit B (red dashed lines).
\label{Fig:tcsc}}
\end{figure}

The total cross sections for $\bar{D} \Lambda_c \to \bar{D} \Lambda_c$,
$\bar{D} \Lambda_c \to \bar{D} \Sigma_c$ and $\bar{D} \Sigma_c \to \bar{D} \Sigma_c$ are shown in Fig.~\ref{Fig:tcsc}.
For both sets of cut-offs, A and B, there is a clearly visible peak in $\bar{D} \Lambda_c \to \bar{D} \Lambda_c$
around 4295~MeV.  A much smaller bump is observed around 4350~MeV.

Those structures can also be seen in the partial-wave amplitudes. The dimensionless partial-wave amplitude $\tau_{\mu\nu}$ is related to the scattering amplitude $T_{\mu\nu}$ from Eq.~(\ref{eq:Tmat}) via a phase factor $\rho$:
\begin{eqnarray}
\tau_{\mu\nu}=-\pi\sqrt{\rho_\mu\rho_\nu}\,T_{\mu\nu}\,,\quad \rho=\frac{k_\mu E_\mu \omega_\mu}{z}\,,
\end{eqnarray}
with $k_\mu$ ($E_\mu$, $\omega_\mu$) the on-shell three-momentum (baryon energy, meson energy) of the initial or final meson-baryon system $\mu$.

\begin{figure}[htbp]
\centering
\includegraphics[width=1\linewidth]{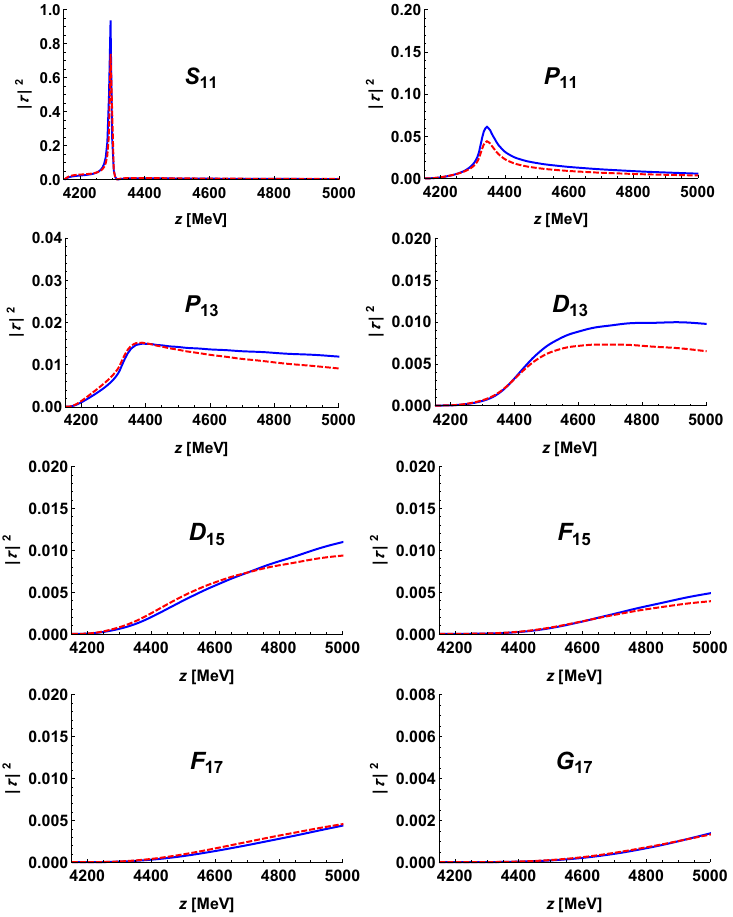}
\caption{Partial-wave amplitudes squared for $\bar{D} \Lambda_c \to \bar{D} \Lambda_c$
using cut-offs A (blue solid lines) and cut-offs B (red dashed lines).
\label{Fig:t2c}}
\end{figure}

In the following, we concentrate on the poles observed in $\bar{D} \Lambda_c \to \bar{D} \Lambda_c$,
and thus only discuss the amplitudes of this process.
The amplitude squared of each partial wave for the $\bar{D} \Lambda_c \to \bar{D} \Lambda_c$ process are shown
in Fig.~\ref{Fig:t2c}. Note that here we only show partial waves up to $J=7/2$ since the higher amplitudes with $J=9/2$ are very flat and smooth.

\begin{figure}[htbp]
\centering
\includegraphics[width=1\linewidth]{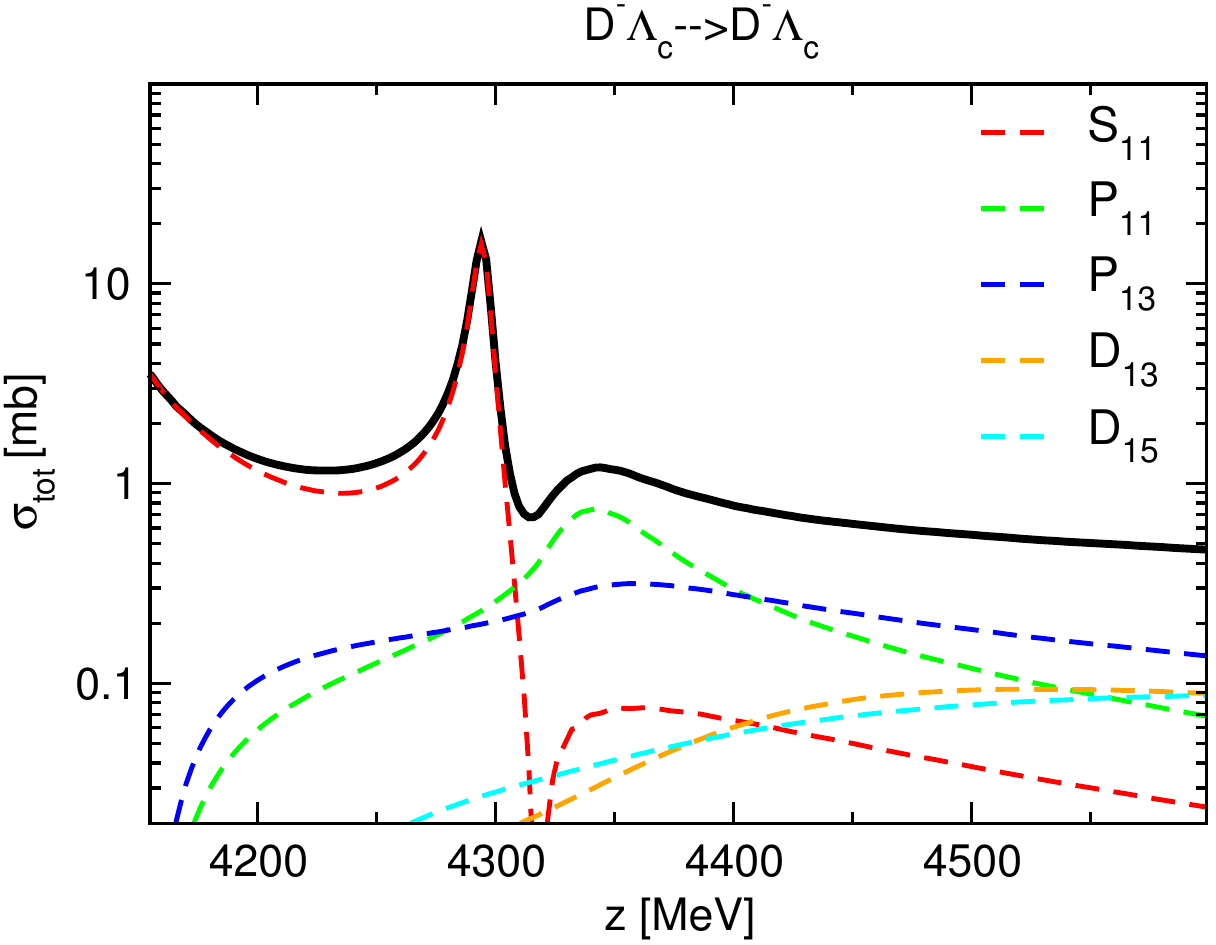}
\caption{Partial-wave content of the total cross section for $D^-\Lambda_c\to D^-\Lambda_c$ using cut-off values A. Black line: full solution. Only dominant partial waves are shown.
\label{Fig:tcsc_pcs}}
\end{figure}

In the $S_{11}$ wave, we can see a very clear sharp peak located around 4295~MeV, while another peak is observed in $P_{11}$ around 4350~MeV. Those structures are responsible for the pronounced peak and the smaller bump in the total cross section of this process. This is reflected in Fig.~\ref{Fig:tcsc_pcs} where the partial-wave content of the total cross section is shown.
Here, one can see that also the structure appearing in $P_{13}$ around 4350~MeV contributes to the smaller bump in the total cross section. In the higher partial waves of $\bar{D} \Lambda_c \to \bar{D} \Lambda_c$, no specific structures are observed and the contribution to the total cross section is small.

\begin{figure}[htbp]
\centering
\includegraphics[width=1\linewidth]{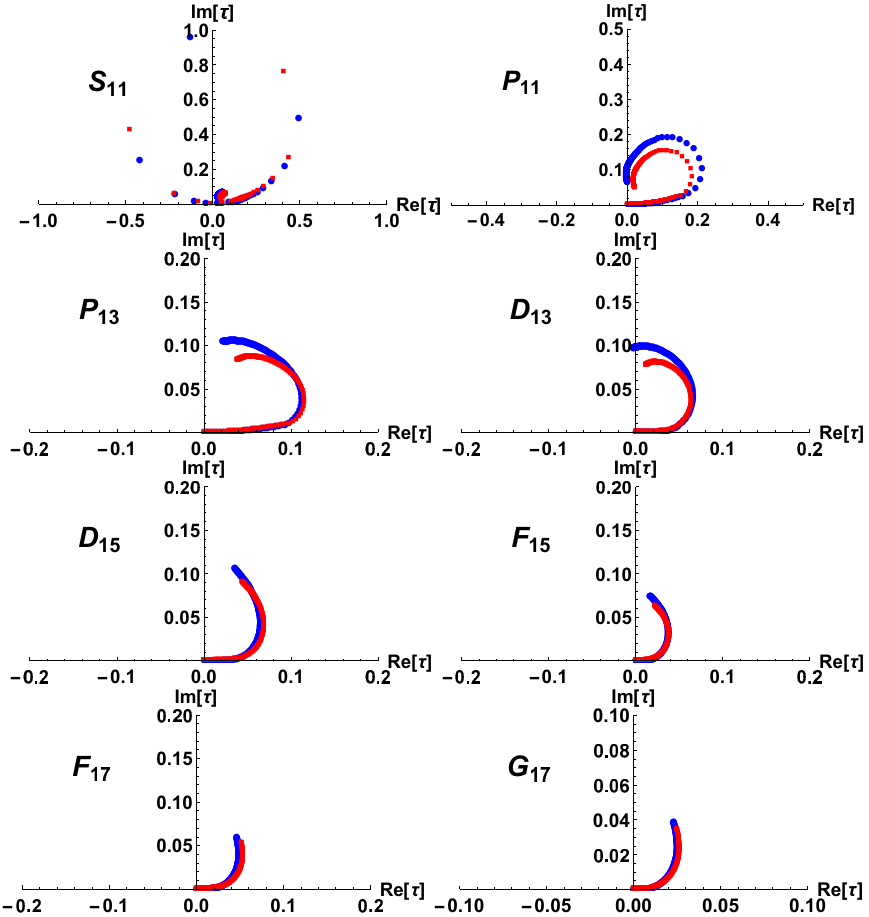}
\caption{Argand plots for different partial waves of $\bar{D} \Lambda_c \to \bar{D} \Lambda_c$
using cut-off values of fit A (blue dots) and of fit B (red squares).
\label{Fig:argc}}
\end{figure}

In addition to the partial-wave amplitudes we also show the Argand diagrams for $\bar{D} \Lambda_c \to \bar{D} \Lambda_c$ in Fig.~\ref{Fig:argc}.
In case of an isolated, Breit-Wigner-type resonance without background contributions,
the Argand plot is a counter-clockwise circle with radius 1/2 whose center and radius can be related to the partial decay width of the state.
However, in a coupled-channel calculation as performed here, with  interference and background effects, the plot will in general not show this behaviour.

In accordance with the sharp peak and almost negligible background  in the $S_{11}$ partial wave in Fig.~\ref{Fig:t2c}, there is a standard full circle in the Argand diagram for $S_{11}$ in Fig.~\ref{Fig:argc}. A clear circular form is also observed for $P_{11}$, while the higher partial waves show no full circles.

The distinctiveness of the peaks in $S_{11}$, $P_{11}$ and, to a certain degree, $P_{13}$ suggests the existence of poles in the scattering matrix, while no conclusion on the resonance content can be drawn from the partial-wave amplitudes with $L\geq 2$ shown in Fig.~\ref{Fig:t2c} or the Argand diagrams in Fig.~\ref{Fig:argc}. In order to substantiate the former hypothesis and clarify the situation in higher partial waves, we perform a pole search in the complex energy plane of the second Riemann sheet of the scattering matrix $T$.  $T_{\mu\nu}^{(2)}$ is accessed via the method of analytic continuation developed in Ref.~\cite{Doring:2009yv}.  We have checked that the pole positions evaluated on the $\bar{D}\Sigma_c$ unphysical sheet are the same as the ones found on the unphysical sheet of the $\bar{D}\Lambda_c$ channel.
The coupling strength $g_\mu$ of a pole at $z=z_R$ to channel $\mu$ is given by the residue $a_{-1,\mu\nu}$ in the Laurent expansion of $T^{(2)}_{\mu\nu}$,
\begin{equation}
T^{(2)}_{\mu\nu}=\frac{a_{-1,\mu\nu}}{z-z_R} + a_{0,\mu\nu}+\mathcal{O}(z-z_R)\,,
\label{eq:laurent}
\end{equation}
with
\begin{equation}
a_{-1,\mu\nu}=g_\mu g_\nu\,.
\label{eq:resid}
\end{equation}
The residues are determined following the procedure explained in Appendix C of Ref.~\cite{Doring:2010ap}.

The pole positions and residues extracted in the present study are listed in table~\ref{table:polec2}. We observe no states with isospin $I=3/2$.
As expected, we find a pole with $J^P=1/2^-$ at $z_R=4295-i\,3.71$~MeV (fit A). As this state is very narrow and located $\sim 25$~MeV below the $\bar D\Sigma_c$ threshold of 4320~MeV, we consider this pole to be a bound state with respect to $\bar D\Sigma_c$. Reducing the comparably large cut-off at the $D\rho D$ vertex results in an even smaller imaginary part until the pole finally moves onto the real axis for $\Lambda_{D\rho D}<1.7$~GeV, while the real part of the pole position moves closer to the $\bar D\Sigma_c$ threshold.

An S-wave bound state was also found in Ref.~\cite{Wu:2010vk} at $z_R=4265-i\,11.6$~MeV, i.e. about 30~MeV lower and 16~MeV broader than in the present study. As in our case, the coupling to $\bar{D} \Sigma_c$ is much stronger than to $\bar{D} \Lambda_c$. It should be noted that in Ref.~\cite{Wu:2010vk} besides $\bar D\Lambda_c$ and $\bar D \Sigma_c$ the $\eta_c N$ channel is included. Including lighter meson-baryon channels the state gains a width of 56.9~MeV in Ref.~\cite{Wu:2010vk}. The existence of an S-wave $\bar D\Sigma_c$ bound state around 4300~MeV was also supported by other dynamical models~\cite{Wang:2011rga,Yang:2011wz,Wu:2012md}.

\begin{figure}[htbp]
\centering
\includegraphics[scale=0.5]{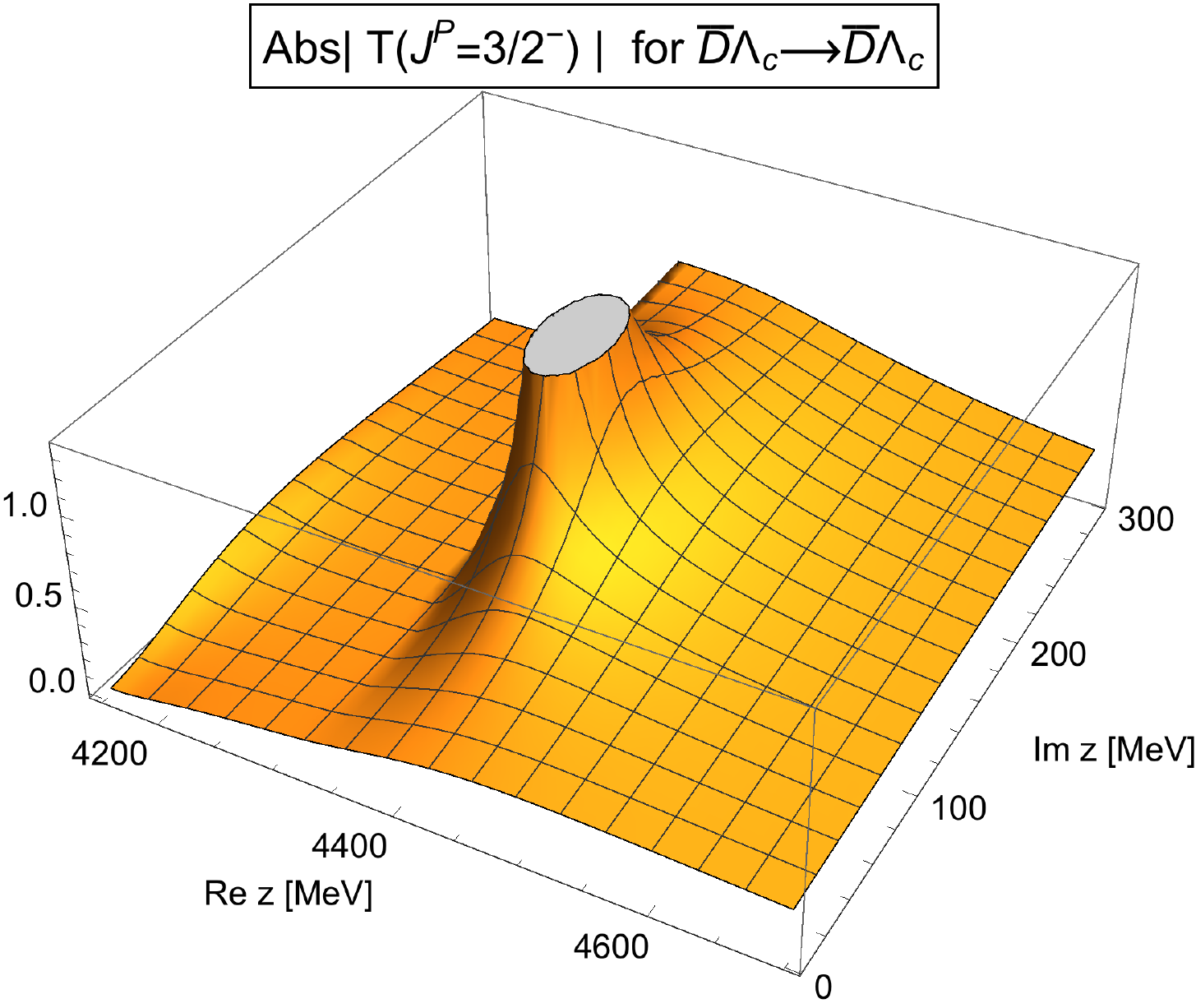}
\caption{Absolute value of $T_{\bar{D} \Lambda_c \, \bar{D} \Lambda_c}$ (second Riemann sheet) with $J^P=3/2^-$. The pole at $z_R=4380-i\,147$~MeV is clearly visible.
\label{Fig:D13}}
\end{figure}

In our present approach, besides the pole in $S_{11}$, one pole is found for $J^P=1/2^+$ and $3/2^+$ each, as suggested by the peaks observed in the partial-wave amplitudes in Fig.~\ref{Fig:t2c}. Both states are located above the $\bar D\Sigma_c$ threshold in the complex energy plane and are, therefore, resonances. In addition we find a resonance state with $J^P=3/2^-$ that does not show up as a peak or bump in $\tau$. In the complex energy plane of $T^{(2)}$, however, the pole is clearly visible, c.f. Fig.~\ref{Fig:D13}. Two more poles with larger imaginary parts not listed in table~\ref{table:polec2} are observed in the $J=5/2$ partial waves.

It can be seen that the results for using the cut-offs from fits A or B are very similar. The values listed in table~\ref{table:polec2} should be considered with care since only two channels are considered in the present exploratory study. The inclusion of $\bar{D}^*\Lambda_c$, $\bar{D}^*\Sigma_c$, $\bar{D}\Sigma_c^*$ and other lighter meson-baryon channels in future studies could have substantial influence on the exact values of the pole positions and residues. This is important when observing that the state found in the $J^P=3/2^-$ partial wave is in agreement with the LHCb $P_c(4380)$.

\begin{table*}[htbp]
    \begin{center}
    \renewcommand{\arraystretch}{1.70}
    \begin{tabular}{c | m{2.5cm} m{2.3cm} m{2.3cm} | m{2.5cm} m{2.3cm} m{2.3cm}}
        \hline
        \hline
        \multirow{4}{*}{$J^P$} & \multicolumn{3}{c|}{A} & \multicolumn{3}{c}{B}  \\
        \cline{2-7}
        & \multirow{2}{*}{$\; z_R$ [MeV]} & \multicolumn{2}{c|}{Couplings [$10^{-3}$ MeV$^{-\frac{1}{2}}$]}
        & \multirow{2}{*}{$\; z_R$ [MeV]} & \multicolumn{2}{c}{Couplings [$10^{-3}$ MeV$^{-\frac{1}{2}}$]}  \\
        \cline{3-4}\cline{6-7}
        &  & $\quad g_{\bar{D}\Lambda_c}$ & $\quad g_{\bar{D}\Sigma_c}$ &  & $\quad g_{\bar{D}\Lambda_c}$ & $\quad g_{\bar{D}\Sigma_c}$  \\
        \hline
         $\frac{1}{2}^-$  &  $4295-i\,3.7$  &  $1.4+i\,0.2$  &  $13.2+i\,0.8$  &  $4297-i\,3.0$  &  $1.1+i\,0.2$  &  $10.9+i\,0.6$  \\
        \hline
        $\frac{1}{2}^+$ & $4334-i\,28$ & $1.1-i\,1.1$ & $-1.9+i\,3.6$ & $4334-i\,30$ & $1.0-i\,1.0$ & $-1.9+i\,3.7$  \\
        \hline
        $\frac{3}{2}^+$ &  $4325-i\,54$ & $0.3-i\,1.1$ & $0.8-i\,4.5$ & $4325-i\,54$ & $0.3-i\,1.0$ &  $0.7-i\,4.6$  \\
        \hline
        $\frac{3}{2}^-$ & $4380-i\,147$ & $0.5-i\,1.9$ & $-1.4+i\,5.6$ & $4378-i146$ & $0.5-i\,1.7$ & $-1.3+i\,5.6$  \\
        \hline
        \hline
    \end{tabular}
    \caption{\label{table:polec2}
    Pole positions $z_R$, couplings and spin-parity for the states in the hidden charm sector with $I=\frac{1}{2}$ using the values of the cut-offs
    for fits A and B.}
    \end{center}
\end{table*}

\subsection{$B \Lambda_b - B \Sigma_b$ interactions}

\begin{figure}[h!]
\centering
\includegraphics[scale=1.3]{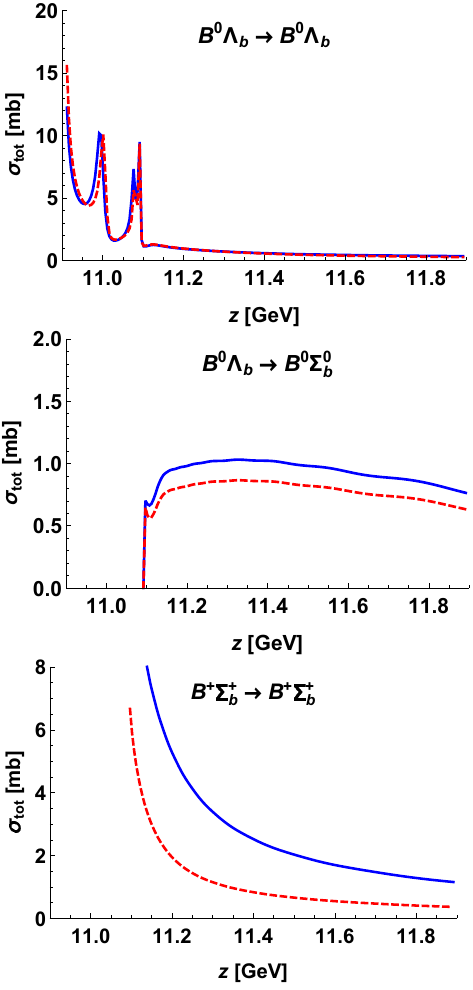}
\caption{Total cross sections for $B \Lambda_b \to B \Lambda_b$, $B \Lambda_b \to B \Sigma_b$
and $B \Sigma_b \to B \Sigma_b$ using cut-offs A (blue solid lines) and cut-offs B (red dashed lines).
\label{Fig:tcsb}}
\end{figure}

The total cross sections for the $B \Lambda_b - B \Sigma_b$ interactions are shown in Fig.~\ref{Fig:tcsb}.
In $B \Lambda_b \to B \Lambda_b$, more structures can be seen than in the total cross section of $\bar{D} \Lambda_c \to \bar{D} \Lambda_c$, which can be explained by the much heavier mass of the $b$-quark compared to the $c$-quark.
One expects that more bound states could be generated and these states should be bound tighter.

\begin{figure}[h!]
\centering
\includegraphics[width=1\linewidth]{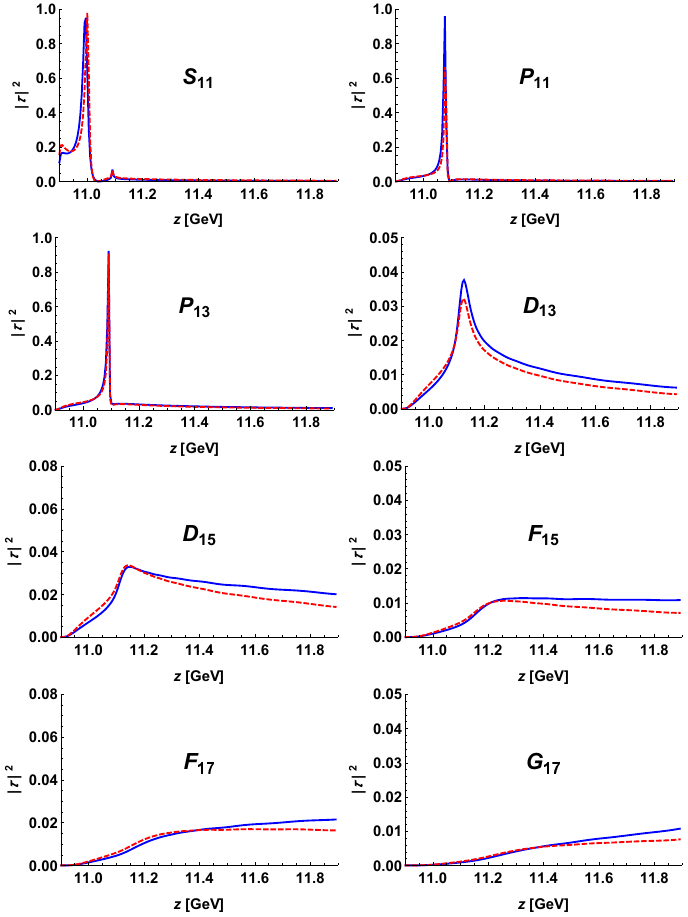}
\caption{Partial-wave amplitudes squared for $B \Lambda_b \to B \Lambda_b$
using the values of the cut-offs for fit  A (blue solid lines) and fit B (red dashed lines).
\label{Fig:t2b}}
\end{figure}

The partial-wave amplitude squared of $B \Lambda_b \to B \Lambda_b$ are shown in Fig.~\ref{Fig:t2b}.
We show partial waves only up to $J=7/2$ as there are no specific structures in higher partial waves.

In $S_{11}$, two peaks appear.
A very distinct structure is observed around 10997~MeV and a much smaller peak is located around 11092~MeV. We consider the latter structure to stem from a kinematical effect at the $B\Sigma_b$ threshold rather than being of dynamical origin. Future studies including additional decay channels and exchange diagrams will help to clarify this issue.
A previous study~\cite{Wu:2010rv} limited to the S-wave approximation also predicts a $B\Sigma_b$ bound state with the binding energy depending on the cut-off parameter for the $t$-channel vector meson exchange, roughly in agreement with present study.

Here, sharp peaks are also observed in $P_{11}$, $P_{13}$ and $D_{13}$,  located around 11075~MeV, 11090~MeV and 11125~MeV, respectively.
Furthermore, there is one possible resonance shape around 11150~MeV in the $D_{15}$ partial wave.

\begin{figure}[h!]
\centering
\includegraphics[width=1\linewidth]{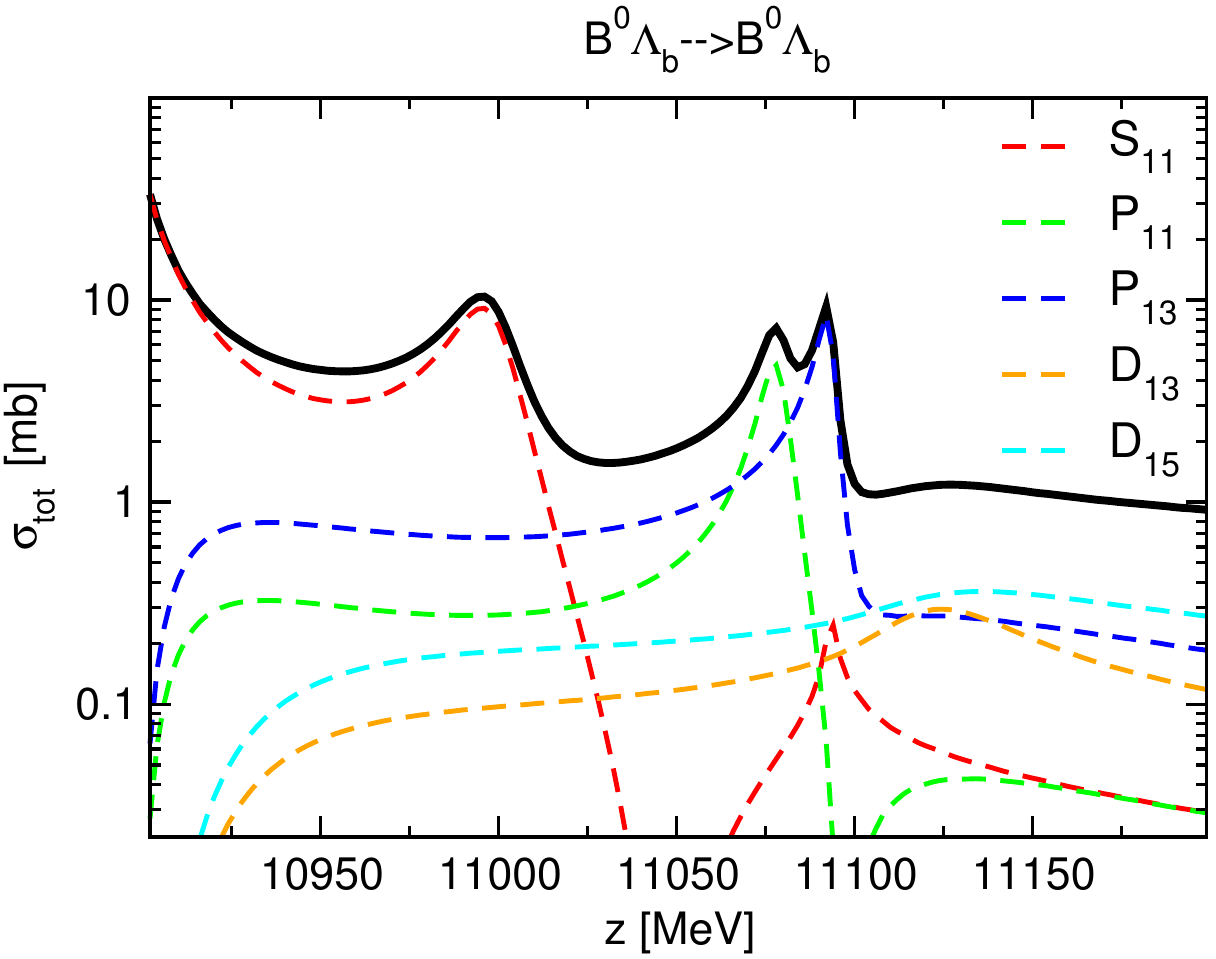}
\caption{Partial-wave content of the total cross section for $B^0\Lambda_b\to B^0\Lambda_b$ using cut-off values A.
Black line: full solution. Only dominant partial waves are shown.
\label{Fig:tcsb_pcs}}
\end{figure}

As can be seen in Fig.~\ref{Fig:tcsb_pcs}, the first sharp peak in the $B \Lambda_b \to B \Lambda_b$ total cross section is caused by the $S_{11}$ partial wave, while the second and third one originate from structures in $P_{11}$ and $P_{13}$. The peaks in the $D$-waves observed in the amplitudes $\tau$ are responsible for a barely visible bump in the total cross section, which can be seen in the logarithmic plot of Fig.~\ref{Fig:tcsb_pcs}.

\begin{figure}[h!]
\centering
\includegraphics[width=1\linewidth]{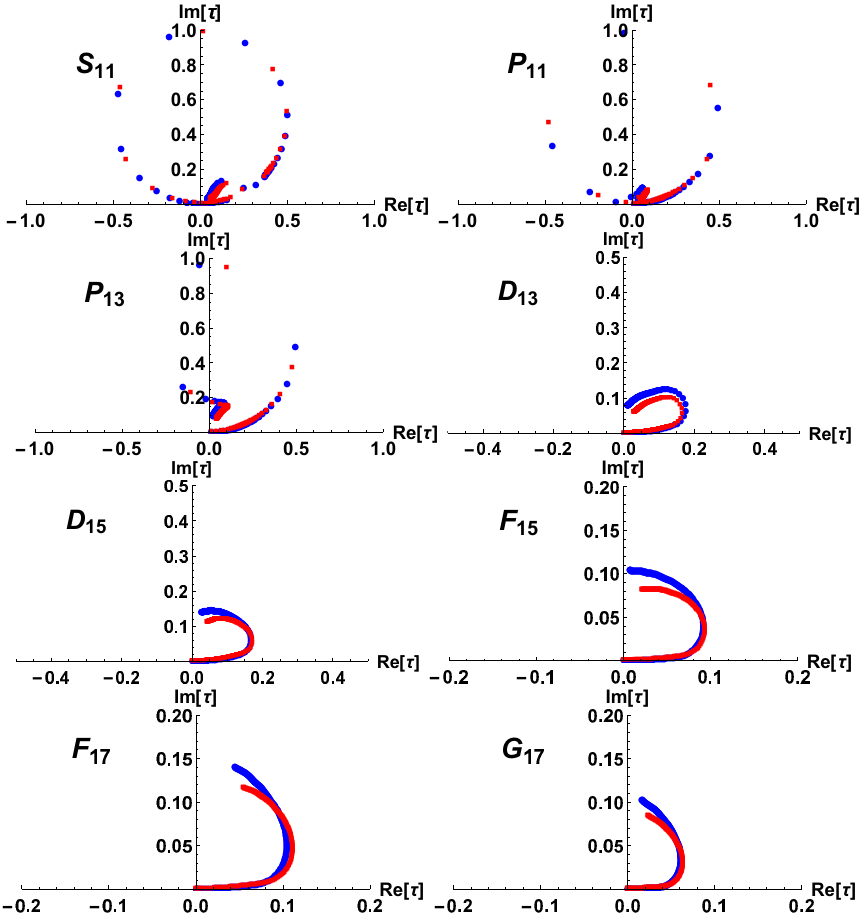}
\caption{Argand plots for different partial waves of $B \Lambda_b \to B \Lambda_b$
using cut-offs A (blue dots) and cut-offs B (red squares).
\label{Fig:argb}}
\end{figure}

The Argand diagrams for the reaction $B \Lambda_b \to B \Lambda_b$ are shown in Fig.~\ref{Fig:argb}. In contrast to the Argand diagrams in the hidden charm sector, there are clear circles with radius 1/2 in several partial waves, namely $S_{11}$, $P_{11}$, and $P_{13}$, which corresponds to a Breit-Wigner-like behaviour and reflects the distinctive peaks  in the partial-wave amplitudes in Fig.~\ref{Fig:t2b}. The $D_{13}$ and $D_{15}$ waves also exhibit a circular form.

\begin{figure}[htbp]
\centering
\includegraphics[scale=0.5]{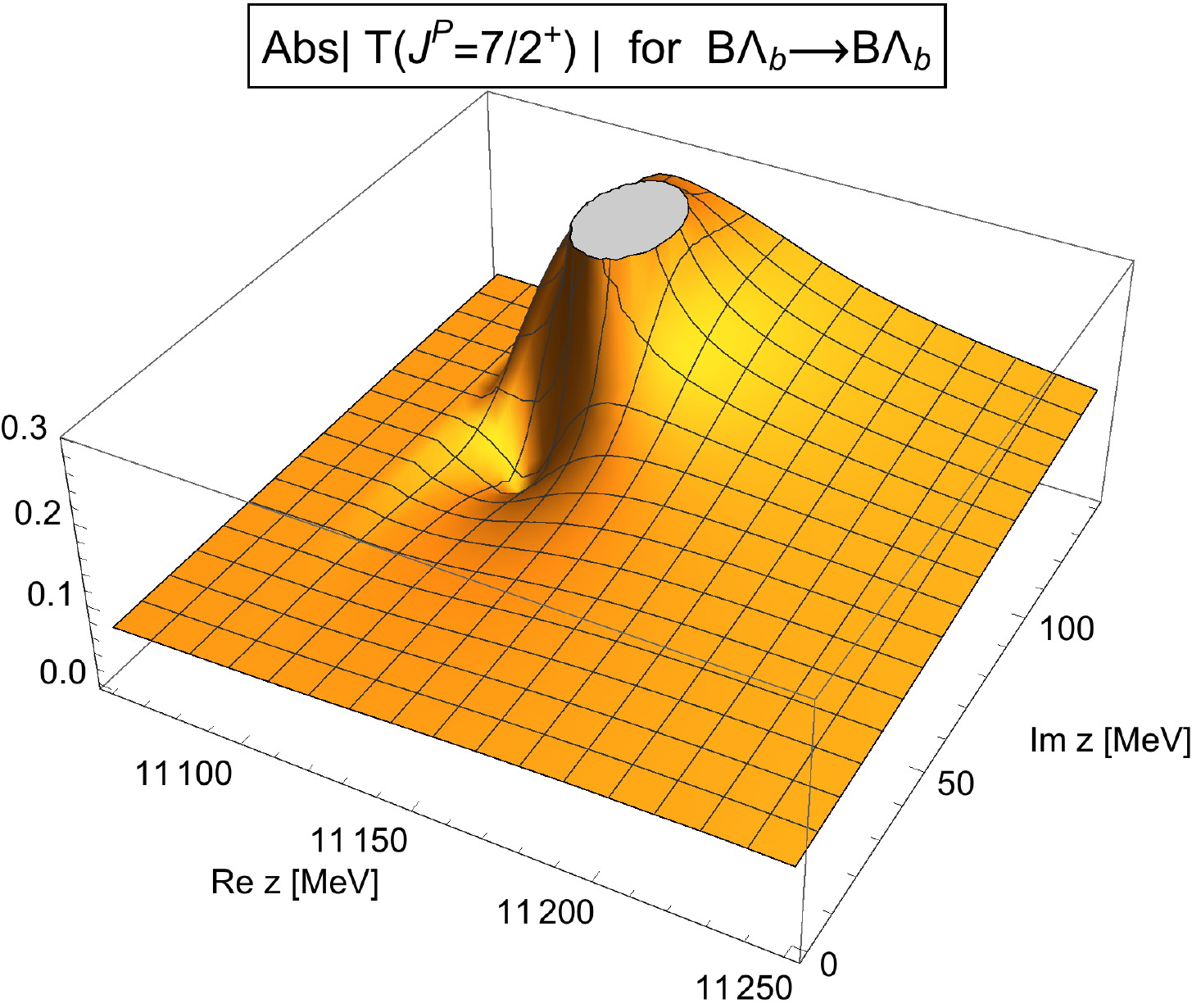}
\caption{Absolute value of $T_{B\Lambda_b \, B\Lambda_b}$ (second Riemann sheet) with $J^P=7/2^+$.
\label{Fig:F17}}
\end{figure}

The pole positions and residues for the $BY_b$ system are given in table~\ref{table:poleb}.
We find two poles below the $B\Sigma_b$ threshold of 11092.84~MeV, one in $S_{11}$, i.e. $J^P=1/2^-$, with a width of about 20~MeV and one narrow state in $P_{11}$, i.e. $J^P=1/2^+$. In the $P_{13}$ wave a very narrow state is found just above the $B\Sigma_b$ threshold. If we reduce the large $B\rho B$ cut-off to 1.7~GeV, this pole moves further away from the threshold and also deeper into the complex energy plane. In case of the narrow state in $P_{11}$, reducing the $B\rho B$ cut-off causes the pole to move towards the $B\Sigma_b$ threshold and onto the real axis. We conclude that the $P_{13}$ state is indeed a resonance, while the state in $P_{11}$ is rather a $B\Sigma_b$ bound state. Further resonances are found for $J^P=3/2^-$, $5/2^+$, $5/2^-$ and also for $J^P=7/2^-$ and $7/2^+$, although for the latter partial waves no resonance structures are observed in the on-shell amplitudes $\tau$ in Fig.~\ref{Fig:t2b}. In contrast, a clear resonance signal can be seen in the complex energy plane of $|T^{(2)}|$, as shown in Fig.~\ref{Fig:F17} for the $F_{17}$ partial wave.

We  remind the reader that, as in case of the hidden charm reactions, the numbers in table~\ref{table:poleb} will change once further decay channels are included in the calculation.

\begin{table*}[htbp]
    \begin{center}
    \renewcommand{\arraystretch}{1.70}
    \begin{tabular}{c | m{2.5cm} m{2.3cm} m{2.3cm} | m{2.5cm} m{2.3cm} m{2.3cm}}
        \hline
        \hline
        \multirow{4}{*}{$J^P$} & \multicolumn{3}{c|}{A} & \multicolumn{3}{c}{B}  \\
        \cline{2-7}
        & \multirow{2}{*}{$\; z_R$ [MeV]} & \multicolumn{2}{c|}{Couplings [$10^{-3}$ MeV$^{-\frac{1}{2}}$]}
        & \multirow{2}{*}{$\; z_R$ [MeV]} & \multicolumn{2}{c}{Couplings [$10^{-3}$ MeV$^{-\frac{1}{2}}$]}  \\
        \cline{3-4}\cline{6-7}
        &  &$\quad g_{B\Lambda_b}$ & $\quad g_{B\Sigma_b}$ &  & $\quad g_{B\Lambda_b}$ &$\quad g_{B\Sigma_b}$ \\
        \hline
         $\frac{1}{2}^-$  &  $   10998-i\,10 $	& $ 1.2+i\,0.3$ & $23.3+i\,1.5 $
         				&  $11005-i\,7$ & $ 1.0+i\,0.3$ & $22.4+i\,1.3$  \\
        \hline
        $\frac{1}{2}^+$ & $11078-i\,4.4  $	&$ 0.7+i\,0.1$ & $-0.9+i\,4.1$
        					& $ 11081-i\,2.8$ & $0.6+i\,0.1$& $-0.6+i\,3.6$  \\
        \hline
        $\frac{3}{2}^+$ &  $  11093-i\,1.8 $	&$0.4-i\,0.004 $ & $0.5-i\,0.7$
        					 & $ 11093-i\,1.5$ & $ 0.4-i\,0.01$ &$0.5-i\,0.7$ \\
        \hline
        $\frac{3}{2}^-$ & $11120-i\,25  $	&$0.3-i\,0.3$ & $-1.1+i\,1.6$
        					& $ 11120-i\,25$ & $ 0.3-i\,0.3$ & $-1.1+i\,1.6$ \\
        \hline
        $\frac{5}{2}^-$ & $11116-i\,38$&$0.2-i\,0.4$ & $ 0.7-i\,2.0$
        					& $ 11116-i\,38$ & $ 0.2-i\,0.4 $ & $  0.7-i\,2.0 $ \\
        \hline
        $\frac{5}{2}^+$ & $11149-i\,88  $	&$ 0.1-i\,0.6 $ & $ -0.8+i\,2.6$
        					& $ 11148-i\,86$ & $ 0.1  -i\,0.5 $ & $-0.8+i\,2.5 $  \\
        \hline
        $\frac{7}{2}^+$ & $ 11123-i\,104 $ &$0.2+i\,0.5$ & $ -0.1+i\,2.6$
        				 	& $ 11124-i\,102$ & $ 0.2+i\,0.5$ & $-0.1+i\,2.5$\\
        \hline
        $\frac{7}{2}^-$ & $ 11176-i\,160  $&$ 0.04+i\,0.3$ & $ 0.4-i\,1.7 $
        					& $ 11171-i\,168$ & $ 0.1+i\,0.5  $ & $0.4-i\,2.3  $ \\
        \hline
        \hline
    \end{tabular}
    \caption{\label{table:poleb}
    Pole positions $z_R$, couplings and spin-parity in the hidden beauty sector with $I=\frac{1}{2}$ using the values for the cut-offs of fits A and B.}
    \end{center}
\end{table*}

\subsection{Summary}

In this exploratory study the J\"ulich-Bonn dynamical coupled-channel model has been extended to the hidden charm and hidden beauty sector. This Lagrangian-based approach respects unitarity and the full off-shell solution of a Lippmann-Schwinger-type scattering equation provides the correct analytic structure. The latter is a prerequisite of a reliable extraction of the baryon spectrum in terms of poles and residues.

As a first step, we included the channels $\bar D\Lambda_c$, $\bar D\Sigma_c$ and $B\Lambda_b$, $B\Sigma_b$ with $\rho$ and $\omega$ exchange in the $t$-channel and, for $\bar DY_c$, $\Xi_{cc}$ exchange in the $u$-channel.
Predictions for cross sections and partial-wave amplitudes are provided and  the possibility of dynamically generated poles has been examined.

In the $\bar{D} \Lambda_c - \bar{D} \Sigma_c$ interactions, we find one pole in each partial wave up to $J^P=5/2^+$ and $5/2^-$. A very narrow pole with $J^P=1/2^-$ is considered to be a bound state with respect to $\bar{D} \Sigma_c$. This bound state was previously predicted in Refs.~\cite{Wu:2010jy,Wu:2010vk}.
In order to obtain a clearer understanding of the nature of these poles and substantiate the pole positions and residues extracted in the present analysis, we need to include further channels like $\bar D^*Y_c$ and other lighter meson-baryon decay modes. Such an analysis is planned for the future. Nevertheless, from our present exploratory study, it is clear that the lowest positive parity P-wave state is only about 40~MeV higher than the lowest negative parity S-wave state. For the compact pentaquark states, the lowest P-wave excitation is predicted to be
more than 130~MeV higher than the lowest S-wave state~\cite{Yuan:2012wz}.

For the $B \Lambda_b - B \Sigma_b$ interactions, we find several poles in partial waves up to $G_{17}$.
Like in the charmed case, more channels should be included to confirm the resonance spectrum extracted here. This task will be addressed in future studies, too.

To achieve a reliable picture of the pentaquark hadron spectrum, it is important to observe the $\bar{D} \Lambda_c$ and $\bar{D} \Sigma_c$ mass spectrum experimentally.\\

\section*{Acknowledgments}

We thank Michael D\"oring, Feng-Kun Guo, and Johann Haidenbauer for useful discussions.
The authors gratefully acknowledge the computing time granted on the HPC Cluster of SKLTP/ITP-CAS and on the supercomputer JURECA at J\"ulich Supercomputing Centre (JSC).
This work is supported in part by DFG and NSFC through funds provided to the Sino-German CRC 110 ``Symmetry and the Emergence of Structure in QCD" (NSFC Grant No. 11621131001, DFG Grant No. TR110), as well as an NSFC fund
under Grant No. ~11647601. 
The work of UGM was also supported by the CAS President's International
Fellowship Initiative (PIFI) (Grant No. 2017VMA0025).

\bibliographystyle{plain}

\begin{thebibliography}{99}

\bibitem{Aaij:2015tga}
  R.~Aaij {\it et al.} [LHCb Collaboration],
  Phys.\ Rev.\ Lett.\  {\bf 115} (2015) 072001
  [arXiv:1507.03414 [hep-ex]].

\bibitem{Shen:2016tzq}
  C.~W.~Shen, F.~K.~Guo, J.~J.~Xie and B.~S.~Zou,
  Nucl.\ Phys.\ A {\bf 954} (2016) 393
  [arXiv:1603.04672 [hep-ph]].

\bibitem{Chen:2015loa}
  R.~Chen, X.~Liu, X.~Q.~Li and S.~L.~Zhu,
  Phys.\ Rev.\ Lett.\  {\bf 115} (2015) 132002
  [arXiv:1507.03704 [hep-ph]].

\bibitem{Chen:2015moa}
  H.~X.~Chen, W.~Chen, X.~Liu, T.~G.~Steele and S.~L.~Zhu,
  Phys.\ Rev.\ Lett.\  {\bf 115} (2015) 172001
  [arXiv:1507.03717 [hep-ph]].


\bibitem{Roca:2015dva}
  L.~Roca, J.~Nieves and E.~Oset,
  Phys.\ Rev.\ D {\bf 92} (2015) 094003
  [arXiv:1507.04249 [hep-ph]].


\bibitem{Mironov:2015ica}
  A.~Mironov and A.~Morozov,
  JETP Lett.\  {\bf 102} (2015) 271
  [arXiv:1507.04694 [hep-ph]].


\bibitem{He:2015cea}
  J.~He,
  Phys.\ Lett.\ B {\bf 753} (2016) 547
  [arXiv:1507.05200 [hep-ph]].


\bibitem{Huang:2015uda}
  H.~Huang, C.~Deng, J.~Ping and F.~Wang,
  Eur.\ Phys.\ J.\ C {\bf 76} (2016) 624
  [arXiv:1510.04648 [hep-ph]].


\bibitem{Burns:2015dwa}
  T.~J.~Burns,
  Eur.\ Phys.\ J.\ A {\bf 51} (2015) 152
  [arXiv:1509.02460 [hep-ph]].

\bibitem{Lu:2016nnt}
  Q.~F.~L\"u and Y.~B.~Dong,
  Phys.\ Rev.\ D {\bf 93} (2016) 074020
  [arXiv:1603.00559 [hep-ph]].

\bibitem{He:2016pfa}
  J.~He,
  Phys.\ Rev.\ D {\bf 95} (2017) 074004
  [arXiv:1607.03223 [hep-ph]].

\bibitem{Lin:2017mtz}
  Y.~H.~Lin, C.~W.~Shen, F.~K.~Guo and B.~S.~Zou,
  Phys.\ Rev.\ D {\bf 95} (2017) 114017
  [arXiv:1703.01045 [hep-ph]].

\bibitem{Guo:2017jvc}
  F.~K.~Guo, C.~Hanhart, U.-G.~Mei\ss ner, Q.~Wang, Q.~Zhao and B.~S.~Zou,
  arXiv:1705.00141 [hep-ph].

\bibitem{Guo:2015umn}
  F.~K.~Guo, U.-G.~Mei\ss ner, W.~Wang and Z.~Yang,
  Phys.\ Rev.\ D {\bf 92} (2015) 071502
  [arXiv:1507.04950 [hep-ph]].


\bibitem{Meissner:2015mza}
  U.-G.~Mei\ss ner and J.~A.~Oller,
  Phys.\ Lett.\ B {\bf 751} (2015) 59
  [arXiv:1507.07478 [hep-ph]].


\bibitem{Liu:2015fea}
  X.~H.~Liu, Q.~Wang and Q.~Zhao,
  Phys.\ Lett.\ B {\bf 757} (2016) 231
  [arXiv:1507.05359 [hep-ph]].

\bibitem{Lebed:2015tna}
  R.~F.~Lebed,
  Phys.\ Lett.\ B {\bf 749} (2015) 454
  [arXiv:1507.05867 [hep-ph]].


\bibitem{Wang:2015epa}
  Z.~G.~Wang,
  Eur.\ Phys.\ J.\ C {\bf 76} (2016)  70
  [arXiv:1508.01468 [hep-ph]].


\bibitem{Scoccola:2015nia}
  N.~N.~Scoccola, D.~O.~Riska and M.~Rho,
  Phys.\ Rev.\ D {\bf 92} (2015) 051501
  [arXiv:1508.01172 [hep-ph]].


\bibitem{Zhu:2015bba}
  R.~Zhu and C.~F.~Qiao,
  arXiv:1510.08693 [hep-ph].

\bibitem{Shimizu:2016rrd}
  Y.~Shimizu, D.~Suenaga and M.~Harada,
  Phys.\ Rev.\ D {\bf 93} (2016) 114003
  [arXiv:1603.02376 [hep-ph]].

\bibitem{Yamaguchi:2016ote}
  Y.~Yamaguchi and E.~Santopinto,
  Phys.\ Rev.\ D {\bf 96} (2017) no.1,  014018
  [arXiv:1606.08330 [hep-ph]].

\bibitem{Wang:2016dzu}
  G.~J.~Wang, R.~Chen, L.~Ma, X.~Liu and S.~L.~Zhu,
  Phys.\ Rev.\ D {\bf 94} (2016) 094018
  [arXiv:1605.01337 [hep-ph]].

\bibitem{Olive:2016xmw}
  C.~Patrignani {\it et al.} [Particle Data Group],
  Chin.\ Phys.\ C {\bf 40} (2016) 100001.

\bibitem{Wu:2010jy}
  J.~J.~Wu, R.~Molina, E.~Oset and B.~S.~Zou,
  Phys.\ Rev.\ Lett.\  {\bf 105} (2010) 232001
  [arXiv:1007.0573 [nucl-th]].

\bibitem{Wu:2010vk}
  J.~J.~Wu, R.~Molina, E.~Oset and B.~S.~Zou,
  Phys.\ Rev.\ C {\bf 84} (2011) 015202
  [arXiv:1011.2399 [nucl-th]].

\bibitem{Wang:2011rga}
  W.~L.~Wang, F.~Huang, Z.~Y.~Zhang and B.~S.~Zou,
  Phys.\ Rev.\ C {\bf 84} (2011) 015203
  [arXiv:1101.0453 [nucl-th]].

\bibitem{Yang:2011wz}
  Z.~C.~Yang, Z.~F.~Sun, J.~He, X.~Liu and S.~L.~Zhu,
  Chin.\ Phys.\ C {\bf 36} (2012) 6
  [arXiv:1105.2901 [hep-ph]].

\bibitem{Yuan:2012wz}
  S.~G.~Yuan, K.~W.~Wei, J.~He, H.~S.~Xu and B.~S.~Zou,
  Eur.\ Phys.\ J.\ A {\bf 48} (2012) 61
  [arXiv:1201.0807 [nucl-th]].

\bibitem{Wu:2012md}
  J.~J.~Wu, T.-S.~H.~Lee and B.~S.~Zou,
  Phys.\ Rev.\ C {\bf 85} (2012) 044002
  [arXiv:1202.1036 [nucl-th]].

\bibitem{Xiao:2013yca}
  C.~W.~Xiao, J.~Nieves and E.~Oset,
  Phys.\ Rev.\ D {\bf 88} (2013) 056012
  [arXiv:1304.5368 [hep-ph]].

\bibitem{Uchino:2015uha}
  T.~Uchino, W.~H.~Liang and E.~Oset,
  Eur.\ Phys.\ J.\ A {\bf 52} (2016),  43
  [arXiv:1504.05726 [hep-ph]].

\bibitem{Gulmez:2016scm}
  D.~G\"ulmez, U.-G.~Mei{\ss}ner and J.~A.~Oller,
  Eur.\ Phys.\ J.\ C {\bf 77}, no. 7, 460 (2017)
  [arXiv:1611.00168 [hep-ph]].

\bibitem{Ronchen:2012eg}
  D.~R\"onchen {\it et al.},
  Eur.\ Phys.\ J.\ A {\bf 49} (2013) 44
  [arXiv:1211.6998 [nucl-th]].

\bibitem{Doring:2010ap}
  M.~D\"oring, C.~Hanhart, F.~Huang, S.~Krewald, U.-G.~Mei\ss ner and D.~R\"onchen,
  Nucl.\ Phys.\ A {\bf 851} (2011) 58
  [arXiv:1009.3781 [nucl-th]].

\bibitem{Ronchen:2014cna}
  D.~R\"onchen {\it et al.},
  Eur.\ Phys.\ J.\ A {\bf 50} (2014),  101
   Erratum: [Eur.\ Phys.\ J.\ A {\bf 51} (2015),  63]
  [arXiv:1401.0634 [nucl-th]].

\bibitem{Ronchen:2015vfa}
  D.~R\"onchen, M.~D\"oring, H.~Haberzettl, J.~Haidenbauer, U.-G.~Mei\ss ner and K.~Nakayama,
  Eur.\ Phys.\ J.\ A {\bf 51} (2015),  70
  [arXiv:1504.01643 [nucl-th]].

\bibitem{Huang:2011as}
  F.~Huang, M.~D\"oring, H.~Haberzettl, J.~Haidenbauer, C.~Hanhart, S.~Krewald, U.-G.~Mei\ss ner and K.~Nakayama,
  Phys.\ Rev.\ C {\bf 85} (2012) 054003
  [arXiv:1110.3833 [nucl-th]].

\bibitem{Doring:2009yv}
  M.~D\"oring, C.~Hanhart, F.~Huang, S.~Krewald and U.-G.~Mei\ss ner,
  Nucl.\ Phys.\ A {\bf 829} (2009) 170
  [arXiv:0903.4337 [nucl-th]].

\bibitem{Schutz:1998jx}
  C.~Sch\"utz, J.~Haidenbauer, J.~Speth and J.~W.~Durso,
  Phys.\ Rev.\ C {\bf 57} (1998) 1464.

\bibitem{Baru:2003qq}
  V.~Baru, J.~Haidenbauer, C.~Hanhart, Y.~Kalashnikova and A.~E.~Kudryavtsev,
  Phys.\ Lett.\ B {\bf 586} (2004) 53
  [hep-ph/0308129].

\bibitem{Kroll:2016mbt}
  P.~Kroll,
  Eur.\ Phys.\ J.\ C {\bf 77} (2017) no.2,  95
  [arXiv:1610.01020 [hep-ph]].

\bibitem{Wu:2010rv}
  J.~J.~Wu, L.~Zhao and B.~S.~Zou,
  Phys.\ Lett.\ B {\bf 709} (2012) 70
  [arXiv:1011.5743 [hep-ph]].

\end{thebibliography}

\end{document}